\documentstyle[psfig,prl,aps]{revtex}

\newcommand{\be}{\begin{equation}}
\newcommand{\ee}{\end{equation}}
\newcommand{\ba}{\begin{eqnarray}}
\newcommand{\ea}{\end{eqnarray}}

\begin{document}
\title{Monopoles in non-Abelian Dirac-Born-Infeld Theory}
\author{N.~Grandi$^a$\, ,
E.F.\ Moreno$^{b\,a}$\thanks{Supported by CUNY Collaborative
Incentive Grant 991999}\thanks{Investigador CONICET, Argentina.}\,
and\,
F.A.\ Schaposnik$^a$\thanks{Investigador CICBA, Argentina}
\\
~\\
{\normalsize\it
$^a$Departamento de F\'\i sica, Universidad Nacional de La Plata}\\
{\normalsize\it
C.C. 67, 1900 La Plata, Argentina}
\\
~\\
{\normalsize\it
$^b$Physics Department, City College of the City University
of New York}\\
{\normalsize\it
New York NY 10031, USA}\\
{\normalsize\it
Physics Department, Baruch College, The City University of
New York}\\
{\normalsize\it
New York NY 10010, USA}}
\date{\today}
\maketitle

\begin{abstract}
We investigate monopole  solutions for the Born-Infeld Higgs
system. We analyze numerically these solutions and compare them
with the standard 't~Hooft-Polyakov monopoles. We also discuss
the existence of a critical value of $\beta$ (the Born-Infeld
``absolute field parameter'') below which no regular solution
exists.
\end{abstract}


\bigskip



\section{Introduction}
Classical solutions for the Dirac-Born-Infeld (DBI) theories  are
useful in the understanding of  brane dynamics
\cite{Pol}-\cite{CM}. In this respect, bions and soliton like
solutions have recently attracted much attention
\cite{G}-\cite{GGT}; in particular, vortex and monopole solutions
have been investigated \cite{G}-\cite{MNS}.

Concerning monopoles, it was shown in \cite{NS2} that it is
possible to construct a DBI action coupled to a Higgs scalar in
such a way that the usual BPS  monopole solution to the
Yang-Mills-Higgs theory also solves the resulting (first-order)
equations of motion. To this end, one has to endow the Higgs
field with dynamics also described by a square-root Born-Infeld
like Lagrangian and also consider the Prassad-Sommerfield
$\lambda \to 0$ limit for the symmetry breaking potential.  Being
the solution that of a  BPS monopole, one does not capture any
features associated with Born-Infeld dynamics and, in particular,
the resulting solution is insensitive to the value of $\beta$,
the ``absolute field'' parameter in Born-Infeld models. In
contrast, a critical value $\beta_c$ was  discovered in a
previous investigation of vortex solutions in Abelian DBI
theories \cite{MNS}, such that no soliton solution exists for
$\beta \leq \beta_c$, this showing how DBI dynamics determines
the nature of soliton solutions.

In this work we shall discuss monopole solutions in $SO(3)$ DBI
gauge theories coupled to a Higgs triplet which enters through the
usual kinetic energy term, $L_{Higgs} \sim  {\rm tr} (D_\mu \phi
D^\mu \phi)$. Concerning the way in which the non-Abelian DBI
scalar Lagrangian is defined, there exist different possibilities
among which we consider taking  (i)  the usual trace over internal
indices  of the square root DBI Lagrangian defined through its
power series expansion and (ii)  the ``symmetric trace''
advocated by Tseytlin \cite{Tse2} as a way to make contact with
the low energy effective action derived from superstring theories.

The paper is organized as follows:  we present in Section II the
$SO(3)$ DBI-Higgs action, discuss the spherically symmetric ansatz
and derive the radial equations of motion both for the usual and
the symmetric trace. In Section III we describe our numerical
solutions and discuss their main properties. We give in Section IV
analytical arguments giving support to the existence of critical
values for $\beta$ bellow which the monopole solution ceases to
exist. Finally we present in Section V a summary of our results
and  the conclusions.

%

\section{The Lagrangian and the Monopole ansatz}

The 't Hooft-Polyakov monopole solution \cite{tH}-\cite{P} to the
equations of motion of the Yang-Mills-Higgs Lagrangian owes its
existence and main properties to the non-Abelian character of an
ansatz for the gauge and scalar fields, mixing space-time and
internal indices in such a way that ensures  topological
non-triviality and regularity of the resulting solution. In order
to look for analogous solutions in the DBI theory, one should
necessarily start from a non-Abelian version of the Born-Infeld
theory and also decide how the Higgs field will be coupled to the
gauge field.

The definition of the DBI theory for a non-Abelian gauge group is
not unique and several alternatives have been discussed in the
literature, \cite{Tse2},  \cite{AN}-\cite{GSS}.  The simplest
extension amounts to define the gauge field Lagrangian in the form
\be
L^{tr}_{DBI} = \beta^2 {\rm tr}  \left( 1 -
\sqrt{1 + \frac{1}{2\beta^2} F_{\mu\nu} F^{\mu \nu}
- \frac{1}{8\beta^4}( F_{\mu \nu}\tilde F^{\mu \nu})^2} \right).
\label{e1}
\ee
Here $F_{\mu \nu}$ is the field strength taking values in the Lie
algebra of the gauge group (which we take for simplicity as
$SO(3)$),
\be
F_{\mu \nu} = \partial_\mu A_\nu -  \partial_\nu A_\mu + e[A_\mu,A_\nu]
\label{2}
\ee
\be
A_\mu = A_\mu^a t^a = \vec A_\mu \cdot \vec t \; ,
\;\;\;\;\;\;\;\;\;\;\;\;
{\vec t} =\frac{\vec \sigma}{\sqrt 2} \; ,
\label{3}
\ee
\be
\tilde F_{\mu \nu} =
\frac{1}{2} \varepsilon_{\mu \nu \alpha \beta}F^{\alpha \beta}
\label{dual}
\ee
and  ``tr'' in eq. (\ref{e1}) represents  the usual trace on
$SO(3)$ indices, with generators normalized so that
\be
{\rm tr} (t^a t^b) = \delta^{ab}
\label{trace}
\ee

A second possibility is to define a symmetric trace operation,
\be
{\rm Str} (t_1,t_2, \ldots , t_N)
\equiv \frac{1}{N!} \sum_\pi {\rm tr} (t_{\pi(1)} t_{\pi(2)} \ldots
t_{\pi(N)})
\label{5}
\ee
with the sum extending over all permutations $\pi$ of the product
of $N$ given  $t's$. Then, the DBI Lagrangian is defined as
\be
L^{Str}_{DBI} = \beta^2 {\rm Str}  \left( 1 -
\sqrt{1 + \frac{1}{2\beta^2} F_{\mu\nu} F^{\mu \nu}
- \frac{1}{8\beta^4}( F_{\mu \nu}\tilde F^{\mu \nu})^2} \,
\right)\; .
\label{sui}
\ee
Remarkably,  the r.h.s. in (\ref{sui}) can be written in this
case in terms of a determinant,
\be
L^{Str}_{DBI}  = \beta^2 {\rm Str}
\left (1-  \sqrt{-{\det} (g_{\mu\nu} + \frac{1}{\beta}  F_{\mu
\nu})} \,\right)
\label{6}
\ee
thus making contact  with the tree level open string effective
action for branes, \cite{Tse2}.  Of course, $g_{\mu \nu}$ in eq.
(\ref{6}) is the $3+1$  usual Minkowski space-time metric, $g_{\mu
\nu} = diag(1,-1,-1,-1)$, and not the pullback of the $d+1$
dimensional Minkowski metric to the $p+1$ dimensional world volume
of the p-brane. It should be mentioned that odd powers of the
field strength $F$ are absent from the expansion of  $ L_{Str} $
this implying that $F$ (although possibly large) should be slow
varying since $ F^3   \sim [D,D]F^2$. In this sense using Str
amounts to some kind of Abelian approximation. It should be noted
that some unsolved problems related to the use of a symmetric
trace have been signaled. They refer to discrepancies between the
results that arise from a symmetrized non-Abelian Born Infeld
theory and the expected spectrum of brane theories \cite{HT}.

Apart from this alternatives related to the way the trace
operation is defined, one has to decide how the Higgs field
dynamics is introduced.  In previous analysis,  DBI monopoles
were constructed by demanding that the {\it usual}
Yang-Mills-Higgs BPS relations also hold in the DBI case
\cite{NS2}.  This amounts to define a Higgs field Lagrangian  in
a Born-Infeld-like way ({\it i.e.,} also under a square root) in
such a way that the model has a  supersymmetric extension
\cite{GGT}, \cite{BP}-\cite{GSS}. Being the BPS relations the
same as in the Yang-Mills-Higgs case, the resulting DBI monopole
solutions are identical to the well-honored Prassad-Sommerfield
exact solutions and have no specific features resulting from the
DBI dynamics. Instead, we shall consider here the usual $SO(3)$
Higgs field Lagrangian and a symmetry breaking potential not
necessarily in the BPS limit. We then propose the following
Lagrangian for the Higgs field:
\be
{\cal L}_{Higgs} =
   \frac{1}{2} D^\mu\vec\phi . D_\mu\vec\phi-V[\phi]
\label{suiss}
\ee
with the scalar triplet  written in the form
\be
\phi = \phi^a t^a = \vec \phi \cdot \vec t \, ,
\label{hi}
\ee
the symmetry breaking potential given by
\begin{equation}
V[\phi] = \frac{\lambda}{4}(\vec\phi \cdot \vec\phi)^2 -\frac{\mu^2}{2}
\vec\phi \cdot \vec\phi \,
\end{equation}
and the covariant derivative defined as
\be
D_\mu \vec  \phi = \partial_\mu \vec \phi + e \vec A_\mu \wedge
\vec\phi \; .
\label{der}
\ee
\subsection*{(i) The equations of motion  for $L^{tr}_{DBI-Higgs}$}
When the trace operation ``tr'' is used, the DBI-Higgs Lagrangian
reads
\be
L^{tr}_{DBI-Higgs}  =  \beta^2 {\rm tr}  \left( 1 -
\sqrt{1 + \frac{1}{2\beta^2} F_{\mu\nu} F^{\mu \nu}
- \frac{1}{8\beta^4}( F_{\mu \nu}\tilde F^{\mu \nu})^2} \right) +
\frac{1}{2} D^\mu\vec\phi \cdot D_\mu\vec\phi-V[\phi]\; .
\label{es1}
\ee
{}From here  on we shall consider purely magnetic configurations
for which $F_{\mu \nu}\tilde F^{\mu \nu} = 0$ . Then the
equations of motion take the form
\be
D^{\mu}\left(   \frac{\vec{F}_{\mu \nu}}{\sqrt{1+\frac{1}{4\beta^{2}}
\vec F^{\mu \nu}\cdot\vec F_{\mu\nu}}} \right)
  = e \vec\phi\times{D_{\nu}\vec{\phi}} \; ,
\label{eq1}
\ee
\be
D^{\mu}D_{\mu}\vec{\phi}=\mu^{2}\vec{\phi} -
\lambda\phi^{2}\vec{\phi}.
\label{eq2}
\ee

We shall consider the usual spherically symmetric 't Hooft-Polyakov
ansatz \cite{tH}-\cite{P},
\begin{eqnarray}
&&\vec{A}_{i}(\vec r)= \frac{K(r)-1}{e}\;
\vec{\Omega} \wedge\partial_{i}{\vec{\Omega}}\; , \label{a}\\
&&\vec A_0(\vec r) = 0 \; , \label{o}\\
&&\vec{\phi}(\vec r)=\frac{H(r)}{er}\;
\vec{\Omega} \; , \label{H}\\
&&\vec \Omega = \vec \Omega(\theta,\varphi) = \frac{1}{r} \vec r\;
,
\label{oo}
\end{eqnarray}
with the appropriate boundary conditions for $K$ and $H$,
\be
\lim_{r \to \infty} K(r) = 0 \, ,
\;\;\;\;\;\;\;\;\;  \lim_{r \to \infty}
\frac{1}{r} H(r) =
\frac{\mu e}{\sqrt \lambda}
\label{K}
\ee
together with the conditions at the origin
\be
K(0) = 1 \, , ~ ~ ~ ~ ~ ~ ~ H(0) = 0 .
\label{HH}
\ee

Inserting ansatz (\ref{a})-(\ref{H}) into the eqs. of motion
(\ref{eq1})-(\ref{eq2}) one gets
\begin{eqnarray}
& &     r^{2}K''- r^{2}\frac{R'}{R} K' =
K\left(
R H^{2} +
K^{2} - 1\right)
\nonumber \\
& & r^2 H''  = 2HK^{2}-\mu^{2}r^{2}
H(1-\frac{\lambda}{e^{2}\mu^{2}r^{2}}H^{2})
\label{radial}
\end{eqnarray}
where
\be
R = \sqrt{ {{1+\frac{1}{\beta^{2}e^{2}r^{4}}
(r^2K'^{2}+\frac{1}{2}(K^{2}-1)^{2})}} }\; .
\label{al}
\ee
It will be convenient to define new dimensionless variables and
parameters,
\begin{eqnarray}
& & \rho = \frac{e\mu r}{\sqrt{\lambda}} \; ,
\nonumber\\
& & \lambda_0 = \lambda / e^2 \; , \nonumber\\
& & \beta_0 = \frac{\beta\lambda}{e\mu^2} \; ,
\label{par}
\end{eqnarray}
so that one finally has
\begin{eqnarray}
& & \rho^{2}K'' = K(RH^{2} + K^{2}-1)+ \rho^{2}\frac{R'}{R}K'\;,
\label{w1}\\
& & \rho^2 H'' =2HK^{2}-\lambda_{0}H(\rho^{2}-H^{2})\;,
\label{w2}\\
& & R =\sqrt{
1+\frac{1}{\beta_0^2 \rho^{4}}(\rho^2 K'^{2}+\frac{1}{2}(K^{2}-1)^{2})
} \; .
\label{final}
\end{eqnarray}

With this ansatz, we can write the energy for the monopole
solution in the form
\be
E = \frac{4\pi \mu}{\sqrt \lambda e}
\int d\rho \rho^2 \left\{
2\beta_0^2(R - 1) + \frac{1}{2\rho^2}
\left(
(H'-\frac{H}{\rho})^2 + \frac{2}{\rho^2} H^2K^2
\right)
+ \frac{\lambda_0}{4}\left(  \frac{H^2}{\rho^2} - 1 \right)^2
\right\}
\ee
This expression reduces to the 't Hooft-Polyakov monopole mass
formula in the $\beta \to \infty$ limit, as expected.

The electromagnetic $U(1)$ field strength ${\cal F}_{\mu \nu}$
is defined as usual \cite{tH} in the form
\be
{\cal F}_{\mu \nu} = \frac{1}{|\phi|} \vec \phi \cdot \vec F_{\mu \nu}
- \frac{1}{e|\phi|^3} \vec \phi \cdot ( {D_\mu \vec \phi} \wedge
{D_\mu \vec\phi}).
\label{F}
\ee
Now, since we are considering DBI dynamics, we have to distinguish
between the magnetic induction $\vec B$ and the magnetic intensity
$\vec H$,
\be
B^i = \frac{1}{2} \varepsilon^{ijk} {\cal F}_{jk} \; ,
\;\;\;\;\;\;\;\;\;\;\;\;\;
H^i = \frac{1}{2} \varepsilon^{ijk} \frac{{\cal F}_{jk} }{R}
\label{BB-HHH}
\ee
Using ansatz   (\ref{a})-(\ref{H}) one easily finds that
\be
B^i  = \frac{x^i}{er^3} \, ,
\label{BBB}
\ee
so that the magnetic flux at infinity,
\be
\Phi \equiv \int_{S^2_\infty} dS_i  B^i = \frac{4\pi}{e} \, ,
\label{fl}
\ee
corresponds to that of a unit magnetic monopole located at the
origin. The magnetic flux $\Phi$ can alternatively be defined in
terms of $\vec H$,  this leading to the same answer (\ref{fl}).
\subsection*{(ii) The equations of motion  for $L^{Str}_{DBI-Higgs}$}
When the symmetric trace operation is used, the DBI-Higgs Lagrangian is
defined as
\be
L^{Str}_{DBI-Higgs}  =  \beta^2\;\; {\rm Str}  \left( 1 -
\sqrt{1 + \frac{1}{2\beta^2} F_{\mu\nu} F^{\mu \nu}
- \frac{1}{8\beta^4}( F_{\mu \nu}\tilde F^{\mu \nu})^2} \right) +
\frac{1}{2} D^\mu\vec\phi D_\mu\vec\phi-V[\phi] \; .
\label{es2}
\ee
Again we will only consider purely magnetic configurations so
$F_{\mu \nu}\tilde F^{\mu \nu} = 0 $. Because of the use of the
symmetric trace, deriving the equations of motion in this case
becomes rather involved. Indeed, one has first to expand  the
square root in $L^{Str}$ in powers of $1/\beta^2$  and at each
order $N$, consider the $N!$ terms which are included in Str. For
example, up to order $1/\beta^2$ one has for the purely DBI
Lagrangian
\be
{\cal L}^{Str} _{DBI}  =  -\frac 1 4 \vec F_{\mu\nu}\cdot\vec
F^{\mu\nu} + \frac 1 {96 \beta^2}
\left(
(\vec F_{\mu\nu}\cdot\vec F^{\mu\nu})^2 + 2(\vec F_{\mu\nu}\cdot\vec
F_{\rho\sigma})(\vec F^{\mu\nu}\cdot\vec F^{\rho\sigma})
\right)
+ O(\frac {F^6}{\beta^4})\; .
\label{order}
\ee
Already at this order, this DBI Lagrangian differs from the one
arising when
one expands $L_{DBI}^{tr}$,
\be
{\cal L}_{DBI}^{tr} =  -\frac 1 4 \vec F_{\mu\nu}
\cdot\vec F^{\mu\nu} +
\frac 1 {32 \beta^2}
(\vec F_{\mu\nu}\cdot\vec F^{\mu\nu})^2 +
O(\frac{F^6}{\beta^4})\; .
\label{or6}
\ee
The $1/\beta^4$ term in the expansion of   $L_{DBI}^{Str}$
involves 120 terms containing  the sixth power of the field
strength,  this making the search of a solution using a numerical
approach too complicated. We  shall here consider  the problem to
the $1/\beta^2$  order  given in equation (\ref{order}) and
analyze how the solution differs from the one obtained using the
more simple ``tr'' operation.

The equations of motion for the gauge field resulting from
(\ref{order}) read
\be
D_\mu \left(
\vec F^{\mu\nu}
-
\frac 1 {12\beta^2}
\left(
(\vec F_{\rho\sigma}.\vec F^{\rho\sigma})\vec F^{\mu\nu}
+
2 (\vec F_{\rho\sigma}.\vec F^{\mu\nu})\vec F^{\rho\sigma}
\right)
\right)
= e \vec\phi\times\vec {D^{\nu}\phi}
\label{sim2}
\ee
while those associated to the Higgs field remain unchanged.

After using the spherically symmetric ansatz
(\ref{a})-(\ref{H}),
eq.(\ref{sim2}) becomes
\begin{eqnarray}
 K''(\rho ) &=& - \left(
       3 K(\rho ) - 6 \rho^4 \beta_0^2 K(\rho ) +
       6 \rho ^4  \beta_0^2 H(\rho )^2 K(\rho ) -
       17 K(\rho )^3 +6 \rho^4 \beta_0^2
        K(\rho )^3  +  45 K(\rho )^5 -\right.\nonumber\\
&&71 K(\rho )^7 + 70 K(\rho )^9 - 42 K(\rho )^{11} +
        14 K(\rho )^{13} - 2 K(\rho )^{15} -
        4 \rho K'(\rho ) +\nonumber\\
&&8\rho K(\rho )^2K'(\rho ) - 4\rho K(\rho )^4K'(\rho ) +
        2\rho^2 K(\rho )^3K'(\rho )^2 -  8 \rho  K'(\rho )^3 -
        4 \rho ^3 K'(\rho )^3 +\nonumber\\
&&32 \rho  K(\rho )^2 K'(\rho )^3 - 48\rho  K(\rho )^4 K'(\rho )^3 +
        32\,\rho \, K(\rho )^6\,K'(\rho )^3 -
        8\,\rho \,K(\rho )^8\,K'(\rho )^3 - \nonumber\\
&&12\,\rho ^2\,K(\rho )\,K'(\rho )^4 +
        36\,\rho ^2\,K(\rho )^3\,K'(\rho )^4 -
        36\,\rho ^2\,K(\rho )^5\,K'(\rho )^4 +\nonumber\\
&&\left. 12\,\rho ^2\,K(\rho )^7\,K'(\rho )^4 -
        8\,\rho ^3\,K'(\rho )^7 -2\rho^2 K(\rho)K'(\rho)^2
        \right) \times \frac{1}{S} \;\;,
\label{larg}
\end{eqnarray}

\begin{eqnarray}
S &=& \rho ^2\,\left( 1 - 6\,\rho ^4\,\beta^2 -
         2\,K(\rho )^2 + K(\rho )^4 + 6\,K'(\rho )^2 +
         6\,\rho ^2\,K'(\rho )^2 - 24\,K(\rho )^2\,K'(\rho )^2 +
         \right. \nonumber\\
&&\left. 36\,K(\rho )^4\,K'(\rho )^2 - 24\,K(\rho)^6\,K'(\rho)^2 +
         6\,K(\rho)^8\,K'(\rho)^2 + 28\,\rho^2\,K'(\rho)^6 \right)
\label{ss}
\end{eqnarray}
while the equation for $H(\rho)$ is still given by equation (\ref{w2}).
Finally, the energy associated to the monopole is given by
\begin{eqnarray}
E &=& \frac{4\pi \mu}{\sqrt{\lambda}e} \int d\rho \left\{ \frac 1
{\rho^2} \left( \rho^2K'^2 + \frac 1 2 ( K^2-1 ) ^2 \right)-
\frac 1 {6 \rho^6\beta_0^2} \left( \left(
\rho^2K'^2 + \frac 1 2 ( K^2-1 ) ^2 \right)^2 +
\right.\right.\nonumber\\
&&\hspace{-1cm}\left.\left.\left( \rho^2K'^4 +
\frac 1 2 ( K^2-1 ) ^4 \right)^2 \right) +
\frac{1}{2} \left( \left( H'-{H\over \rho} \right) ^2
-{{2H^2K^2}\over {\rho^2}} \right) - \frac {\lambda_0} 4
\left( \frac{H^2}{\rho^2} - 1 \right)^2 \right\}   +
{\rm O}\!\left( \beta_0^4 \right) \; .
\label{ener2}
\end{eqnarray}
%


\section{Numerical results}
To obtain a detailed profile of the monopole solution, we solved
numerically the differential equations (\ref{w1})-(\ref{final})
for the case of the trace operation ``tr'' and (\ref{larg}),
(\ref{ss}), and (\ref{w2}) for the symmetric trace ``Str''.  We
employed a relaxation method for boundary value problems
\cite{NR}.  Such method determines the solution by starting with
an initial guess and improving it iteratively.  The natural
initial guess was the exact Prassad-Sommerfield solution \cite{PS}
(which corresponds to
$\lambda_0 = 0$ and $\beta \to \infty$).%

\subsection*{(i) The usual trace}

\begin{figure}
\centerline{ \psfig{figure=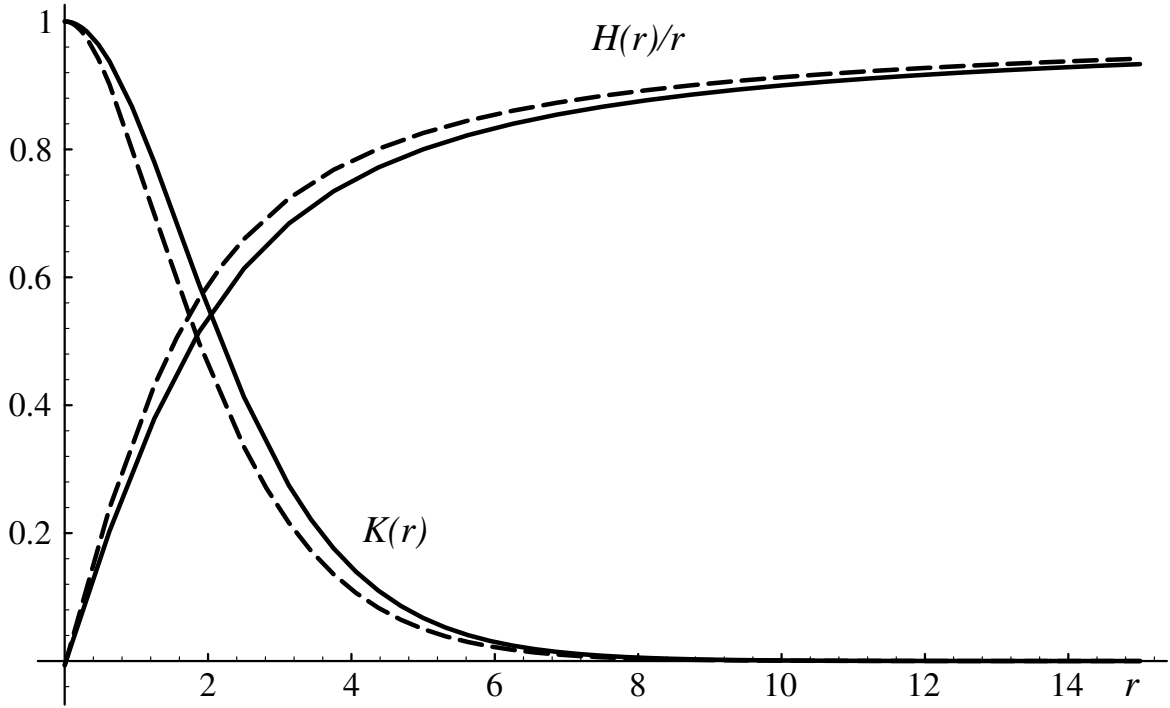,height=20cm,angle=0}}
\smallskip
\caption{ Plot of the functions $K(r)$ and $H(r)/r$ (in
dimensionless variables) for the monopole solution with
$\lambda=0$. The solid line corresponds to the solution with
$\beta=10$ and the dashed line corresponds to the the solution
with $\beta=0.5$. \label{fig-lambda0} }
\end{figure}

\begin{figure}
\centerline{ \psfig{figure=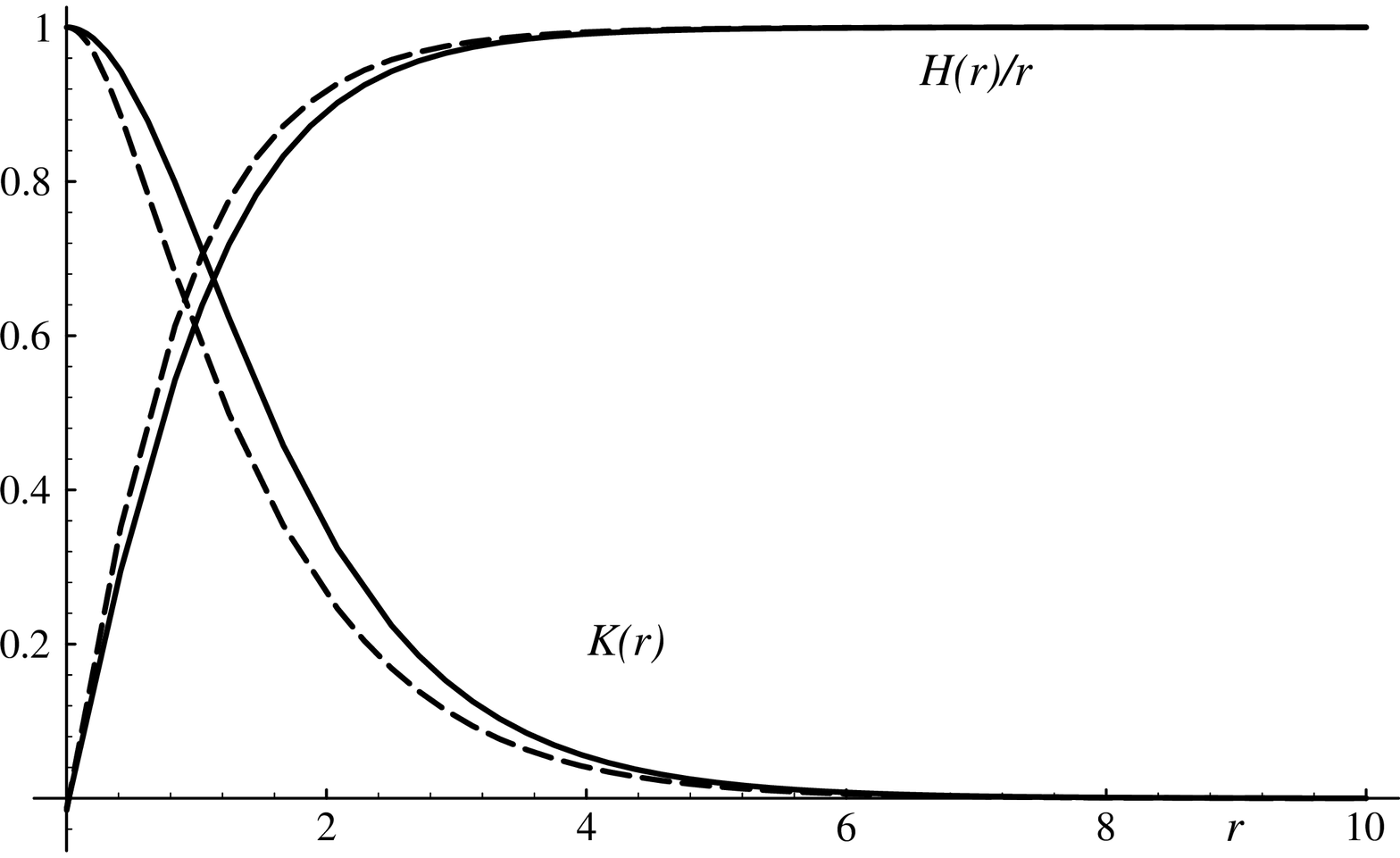,height=20cm,angle=0}}
\smallskip
\caption{ Plot of the functions $K(r)$ and $H(r)/r$ (in
dimensionless variables) for the monopole solution with
$\lambda=0.5$. The solid line corresponds to the solution with
$\beta=10$ and the dashed line corresponds to the the solution
with $\beta=0.8$. \label{fig-lambda05} }
\end{figure}

For $\beta {\
\lower-1.2pt\vbox{\hbox{\rlap{$>$}\lower5pt\vbox{\hbox{$\sim$}}}}\
} 10$, the solutions to eqs.  (\ref{w1})-(\ref{final}) do not
differ appreciably from the 't~Hooft-Polyakov monopole solution
(see for example \cite{JZ} for a plot of the 't~Hooft-Polyakov
solution). As $\beta$ decreases, the solution changes slowly: the
monopole radius decreases and the (radial) magnetic field $\vec H$
concentrates at the origin. Some of the solutions profile are
depicted in figures (\ref{fig-lambda0}) and (\ref{fig-lambda05}).


For $\beta \sim 1$ new features become apparent from our
numerical analysis. In particular, we found that:

~

\noindent  1- For $\lambda_0 = 0$ the behavior of the Higgs field
at large distances
depends on $\beta$,
\be
H(r) \to \phi_0 r + c(\beta/\mu^2) ~ ~ {\rm for} ~ ~ r  \to \infty
\label{as}
\ee
(here $\phi_0$ is the Higgs field v.e.v. in the
Prassad-Sommerfield limit). Of course, for $\beta \to \infty$,
$c(\beta/\mu^2 ) \to 1$  and one has the usual asymptotic behavior
of the Higgs field for the 't Hooft-Polyakov monopole. For finite
$\beta$, however, it is interesting to note that the $1/r$
falloff, which  in the Yang-Mills-Higgs case is related to the
massless dilaton associated to the scale invariance of the BPS
regime,  has now  a $\beta$ dependent coefficient. ~

\noindent 2- There is a critical value of $\beta$, which we shall
denote as $\beta_c$, such that for $\beta \leq \beta_c$ there is
no (numerical) solution to the equations of motion
(\ref{w1})-(\ref{final}). Some values for $\beta_c$  are: $\beta_c
= 0.41$ for $\lambda_0 = 0$,  $\beta_c = 0.62$ for $\lambda_0 =
0.5$.

\begin{figure}
\centerline{ \psfig{figure=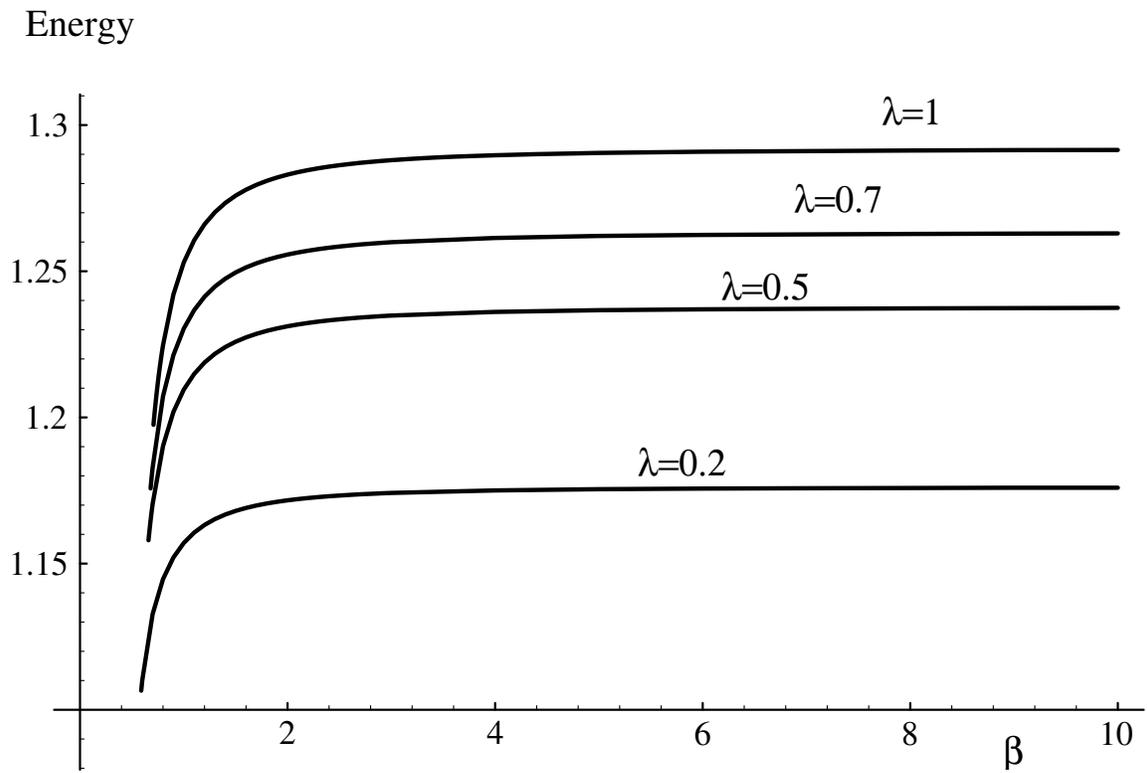,height=20cm,angle=0}}
\smallskip
\caption{ Energy of the monopole configuration as a function of
$\beta$ for different values of $\lambda$. \label{fig-energy} }
\end{figure}

This peculiar characteristic of the solutions  appears to be a
consequence of the high nonlinearity of the equations and not a
fictitious artifact of the numerical method. Moreover, the energy
of the solutions seems to be singular at $\beta=\beta_c$ (see
figure (\ref{fig-energy})).

\subsection*{(ii) The symmetric trace}

For this case we solved the eqs. (\ref{larg}), (\ref{ss}) and
(\ref{w2}) with the same numerical approach. Since the equations
are valid to order $1/\beta^2$,  our analysis cannot be reliable
for small $\beta$. We see that for $\beta  {\
\lower-1.2pt\vbox{\hbox{\rlap{$>$}\lower5pt\vbox{\hbox{$\sim$}}}}\
} 4 $ the solutions do not differ notably from those arising when
the trace operation ``tr'' is considered. The profile of the
solutions are indistinguishable from the solid-line curves of
figures (\ref{fig-lambda0}) and (\ref{fig-lambda05}). In view of
our approximation, we could not analyze the region where one
expects the existence of $\beta_c$ in this case.


\section{Analysis of $\beta_c$}
As noted in the introduction, the existence of a critical value
of the absolute field below which the solution to the equations
of motion of the DBI-Higgs system ceases to exist,  was already
noticed for vortices in the Abelian case \cite{MNS} and should
be considered as a distinctive feature of soliton solutions in
DBI theories.

In order to better understand the origin of  $\beta_c$ let us
introduce the following scaling argument. For a
monopole like solution, there is a characteristic radius $R_W$
that can be associated with the monopole core; outside this
core, the gauge field approaches to its asymptotic value.  For
Yang-Mills-Higgs  monopoles (or Nielsen-Olesen vortices), this
radius should be necessarily related to $\mu$, the sole parameter
carrying dimensions, $R_W \sim 1/\mu$.   The size $R_W$ is fixed
so as to minimize the sum of the energy stored in the magnetic
field outside the core and the energy due to the scalar field
gradient inside the core. The resulting value $R_W$ is in this
case $R_W = 1/M_W$, with $M_W$ the mass of the gauge boson, $M_W=
(e/\sqrt{\lambda} )\mu$. A second length playing a r\^ole in the
monopole configuration is related to the size of the region
outside of which the Higgs scalar takes practically its vacuum
expectation value.  We shall call the radius of this region
$R_H$. For the Yang-Mills-Higgs system one has $R_H = 1/\mu$.
From $R_W$ and $R_H$ we can define a dimensionless parameter $ v$
measuring the relative intensity of the two coupling constants in
the theory,
\be
v \equiv \frac{R_H}{R_W} = \frac{e}{\sqrt \lambda}
\label{v}
\ee
(In the vortex case $1/v$ coincides with the Ginzburg-Landau
parameter separating the two types of superconductivity). For $v
\sim 1$ one has a well defined monopole configuration.

Now, when DBI-Higgs monopoles are considered, there is, apart
from   $\mu$,  a second dimensionful parameter,  $\beta$,
$[\beta] = [\mu]^2$.  Then $R_W$  and $R_H$ could in principle
depend both on $\mu$ and $\beta$ and  the configuration
minimizing the energy  will result from the  matching of both
parameters determining the size of  the monopole.  It may happen
that in some region of the $(\beta,\mu)$ domain such a matching
becomes not possible. The outcome will be the non-existence of
solutions in a range of values of $\beta$ with  size related to
$\beta_c$. In view of the complexity of the non-linear coupled
system (\ref{w1})-(\ref{final}), let us analyze this possibility
by using an approximate monopole configuration sharing the main
features of the true solution:
\be
K_{app}(r) = \left(1-\frac r R\right)
\theta(R-r)
\label{apk}
\ee
\be
H_{app}(r)=r\left(1- \frac{r_0^2}{r^2}\right)^2
\theta(r-r_0).
\label{aph}
\ee
Here $R$  and  $r_0$ are parameters controlling the shape of the
gauge field and scalar field configurations and they have to be
determined by minimizing the energy $E$ of the configuration. One
can relate $R$ and $r_0$ with $R_W$ and $R_H$ by searching in
(\ref{apk})-(\ref{aph}) for the values of $r$ for which the gauge
and scalar field configurations differ in $\frac{1}{\rm e}$ from
its asymptotic. This gives $R_W \sim (1/2)  R $, $R_H \sim 3  r_0
$ and in this sense one can think that $R = R(\beta,\mu)$ and
$r_0 = r_0(\beta,\mu)$. Let us finally note that using
eq.(\ref{v}) one can write $v$ in terms of $r_0$ and $R$,
\be
v \sim 6 \frac{r_0}{R}  \equiv 6 x \; .
\label{vv}
\ee

We have seen that for large $\beta$ (say $\beta  {\
\lower-1.2pt\vbox{\hbox{\rlap{$>$}\lower5pt\vbox{\hbox{$\sim$}}}}\
}5$),  the DBI-Higgs  theory just gives the same answer as the
Yang-Mills-Higgs model so that one should always find in this
region  values for $R$ and $r_0$ minimizing $E$.  In particular,
in the $\beta \to \infty$ case we found using our approximate
configurations that one has, for $\lambda_0 = 0.5$,    $x =
0.128$. This giving for  $v$ the result approximate result
$v_{app} \sim  6 x \sim 0.8$ to be compared with the ``exact''
result  for Yang-Mills-Higgs theory, $v =
1/\sqrt{\lambda_0} = \sqrt{2}$.%

Now, for small $\beta$ the situation radically changes.  Indeed, using
(\ref{apk}) and (\ref{aph}) one finds
for the energy, to second order in $\beta$ (apart of an
irrelevant additive constant):
\begin{eqnarray}
E &=& -R \left( -0.33 + 0.03 {\it \beta} + 3.04\,x + 3.33\,{x^2}
- 8.53\,{x^3} + 2\,{x^4} -\right.\nonumber\\
&&\;\;\;\;\;\;\;\;\;\;\;\;\;\;\;\;\;\;\;\;\;\;\;\;\;\;\;\;\;\;\;\;\;
 \left. 0.13\,{x^6} + 0.01\,{x^8} +
8\,{x^2}\,\log (x) \right)+ 2.1 R^3
{\it \lambda}{x^3} + {\rm O}(\beta^3) \nonumber\\
&=& 2.1 R^3\lambda x^3 - R\left( f(x) + 0.03 \beta \right)
+ {\it O}(\beta^3)\; .
\label{SS}
\end{eqnarray}

It is not difficult to show that for small $\beta$ (and any
$\lambda$) the above expression does not have a minimun for any
$R$ and $x$.
%
%
We conjecture that this phenomenon occurring for an approximate
configuration also takes place for the actual monopole solution:
below a critical $\beta$ value, there is no possible matching
between the monopole core and the size of the region where the
Higgs scalar is different from its vacuum value, in such a way the
energy is minimized.


\section{Summary and conclusions}
We have discussed in this work monopole solutions for an $SO(3)$
Dirac-Born-Infeld gauge theory coupled to a Higgs scalar.  We
considered two alternative Lagrangians for the theory, differing
in the way the trace over group indices is taken. Concerning the
Higgs field, we have chosen the usual kinetic energy term and
symmetry breaking potential.

As in the case of the Yang-Mills-Higgs system, a spherically
symmetric ansatz leads to a system of coupled non-linear radial
equations that have to be solved numerically. We have seen that
the magnetic field corresponds, as in the 't~Hooft-Polyakov case,
to that of a monopole with unit charge. When the absolute field
$\beta$ parameter is large
(${\
\lower-1.2pt\vbox{\hbox{\rlap{$>$}\lower5pt\vbox{\hbox{$\sim$}}}}\ }
5$)
the profile of the monopole solution is practically the same
as the corresponding to the Yang-Mills-Higgs model.  As $\beta$
decreases, the monopole radius becomes smaller and the magnetic
field concentrates more and more near the origin.

A remarkable effect takes place for small $\beta$: there exists a
critical value $\beta_c$   such that for $\beta \leq \beta_c$ the
solution ceases to exist.  The actual value of $\beta_c$ depends
on the choice of the other free parameters. We presented an
scaling argument that supports this result: using an approximate
solution that depends only on the dimensions of the configuration
we showed that for small values of $\beta$ it is impossible to
adjust the size parameters to minimize the energy.


The monopole solution we have presented has many remarkable
features that make worth a thorough investigation. In particular,
the analysis of  dyon solutions, which implies the inclusion of
the $(F\tilde F)^2$ term in the DBI action should reveal new
features related to the existence of the dimensionfull parameter
$\beta$. We hope to discuss this problem in a future work.



\underline{Acknowledgements}: E.~F.~M. wish to thanks V.~P.~Nair for
suggesting the scaling argument of section IV. F.~A.~S. is
partially  supported by CICBA, CONICET (PIP 4330/96), ANPCYT
(PICT 97/2285) Fundaci\'on Antorchas, Argentina and a Commission
of the European Communities contract No:C11*-CT93-0315. This work
is supported in part by funds provided by the U.S. Department of
Energy (D.O.E.) under cooperative research agreement \#
DF-FC02-94ER40818.

\end{document}